\documentclass[prd,10pt,twocolumn,superscriptaddress,floatfix,showpacs,nofootinbib]{revtex4-1}
\usepackage[utf8]{inputenc}  %Input what you want e.g., é, ł, a, ü
\usepackage[T1]{fontenc}     %Output what you want e.g., é, ł, a, ü
\usepackage[british]{babel}  %Do hyphenation according to british english
\usepackage[sc,osf]{mathpazo}
\usepackage[scaled=0.86]{berasans}  % URL font that go well wtih palatino
\usepackage[colorlinks=true, citecolor=blue, urlcolor=blue]{hyperref} 
\usepackage{graphicx} % Package to insert exteral figures
\usepackage[babel]{microtype}  %Improves text justification
\usepackage{amsmath,amssymb,amsthm,bm,amsfonts,mathrsfs,bbm} %Usefull math packages
\usepackage{xspace}  %Useful to add space in macros
\usepackage{pgfplots}
\definecolor{dblue}{rgb}{0,0,0.6}
\definecolor{dred}{rgb}{1,0.08,0.58}

\begin{document}
\title{\textcolor{dblue}{Natural $\alpha$-Attractors from ${\cal N}=1$ Supergravity via flat K\"ahler Manifolds }}

%for Single FieldInflation via ${\cal N}=1$ Supergravity without No-scale%
\author{Tony Pinhero}
\email{tonypinheiro2222@gmail.com}
\affiliation{Physics and Applied Mathematics Unit, Indian Statistical Institute, 203 B. T. Road, Kolkata 700108, India.}

%%%%%%%%%%%%%%%%%%%%%%%%%%%%%%%%%%%%%%%%%%%%%%%%%%%%%%%%%%%%%%%%%%%%%%%%%%%%%%%%%%%%%%%%%%%%%%%%
%%%%%%%%%%%%%%%%%%%%%%%%%%%%%%%%%%%%%%%%%%%%%%%%%%%%%%%%%%%%%%%%%%%%%%%%%%%%%%%%%%%%%%%%%%%%%%%%

\begin{abstract}
We present $\alpha$-attractor models for inflation based on ${\cal N}=1$ supergravity with flat K\"ahler manifolds. The function form of the associated K\"ahler potential in these models are logarithmic square in nature and has a visible shift symmetry in its composite canonical variables. The scalar potential $V$ with respect to these field variables has an infinitely long dS valley of constant depth and width at large values of inflaton field $\psi$ and attains a Minkowski minimum at small $\psi$. We illustrate this new framework with a couple of examples.
\end{abstract}
\pacs{ }
\maketitle 
%%%%%%%%%%%%%%%%%%%%%%%%%%%%%%%%%%%%%%%%%%%%%%%%%%%%%%%%%%%%%%%%%%%%%%%%%%%%%%%%%%%%%%%%%%%%%%%%
\section{Introduction}\label{sec_intro}
One can give the best definition to cosmological $\alpha$-attractors \cite{kallosh2013universality, kallosh2013multi,kallosh_nonminimal_attra2013,kallosh2013sup_alpha_attra,unity_of_cosmo_attracts,escher_in_sky,single_field_andre_linde,hyperbolic_geometry_of_attrctors,seven_disc_manifold,b_mode,ana_achucarro_multi-field_alph_attra,hypernatural_inflation,scalisi_alpha_scale,scalisi_desitter_landscape,pole_nflation} in terms of the non-canonical real field variable $\phi$, through a toy Lagrangian as,
\begin{equation}\label{alpha_attractor}
L=\sqrt{-g}\left[\frac{1}{2}R-\frac{1}{\left(1-\frac{\phi^{2}}{6\alpha}\right)^{2}}\frac{1}{2}\partial_{\mu}\phi\partial^{\mu}\phi-V\left(\phi\right)\right]
\end{equation}
One of the key properties of this theory is that, inflationary observational predictions are to a large extent determined by the field space metric or simply, by the geometry of the moduli space, but not by the potential term. This means that in the leading order approximation in $1/N_{e}$, where $N_{e}$ is the number of e-folds, the observational predictions are stable under significant modifications of the inflationary potential. The reason for such observational predictions of these theories is that their kinetic term has a second order pole. To understand this in detail, we provide a brief technical discussion of these models \cite{unity_of_cosmo_attracts}. Models of the form defined in (\ref{alpha_attractor}) can be generally written in terms of the field variable $\rho$ as,
\begin{equation}
L=\sqrt{-g}\left[\frac{1}{2}R-\frac{1}{2}K_{E}(\rho)\left(\partial\rho\right)^{2}-V_{E}(\rho)\right]
\end{equation}
Now, if we assume that the pole of the kinetic term is located at $\rho=0$, the Laurent expansion of the $K_{E}$ term is as follows
\begin{equation}
K_{E}=\frac{a_{p}}{\rho^{p}}+\dots,~~~~~~~~~~~~V_{E}=V_{0}\left(1+c\rho+\dots\right)
\end{equation}
Where $a_{p}$ is the residue at the leading order pole of order $p$. With the use of these information, one can derive the expression for the spectral index $n_{s}$ and tensor-to-scalar ratio in the leading order approximation of $1/N_{e}$ \cite{unity_of_cosmo_attracts}:
\begin{equation}\label{pole_predictions}
n_{s}=1-\frac{p}{p-1}\frac{1}{N_{e}},~~~~~~~r=\frac{8c^{\frac{p-2}{p-1}}a_{p}^{\frac{1}{p-1}}}{(p-1)^{\frac{p}{p-1}}}\frac{1}{N_{e}^{\frac{p}{p-1}}}
\end{equation}
In this expression, note that the spectral index completely depends upon the order of the pole whereas the tensor-to-scalar ratio depends not only on the leading pole and its corresponding residue. One can see that the contribution from the potential term $c$ vanishes for the pole of order two in the expression of $r$.  The Lagrangian defined in (\ref{alpha_attractor}) has a leading pole of order two with residue $\frac{3}{2}\alpha$, and hence the observational predictions are insensitive to the structure of the potential. As a result, (\ref{pole_predictions}) boils down to
\begin{equation}\label{observational_predictions}
n_{s}=1-\frac{2}{N_{e}}~~~~~~~~~~~~~r=\frac{12\alpha}{N_{e}^{2}}
\end{equation}
for the theory (\ref{alpha_attractor}). These predictions are exactly matching with the latest observations made by Planck \cite{planck2015_inflation,planck2018_inflation}.

It is well-known that these models (\ref{alpha_attractor}) can be successfully embedded in different Supergravity theories. In particular, the following equivalent choices of the K\"ahler potential \cite{kallosh2013sup_alpha_attra,escher_in_sky}
\begin{equation}
K=-3\alpha\log\left(1-ZZ^{*}\right)~~~\text{or}~~K=-3\alpha\log\left(T+T^{*}\right),
\end{equation}
along with different choices of superpotentials embeds these models in $\mathcal{N}=1$ Supergravity. Here $Z$ and $T$ are the inflaton superfields in terms of the disk variables and half plane variables respectively. The particular choice of the superpotential depends on whether the stabilizer field $S$ is inside or outside the logarithmic term of the K\"ahler potential. Such an embedding is also possible with slight modifications in the above defined K\"ahler potentials (see \cite{hyperbolic_geometry_of_attrctors,scalisi_alpha_scale,single_field_andre_linde}). All these models are based on the logarithmic K\"{a}hler potentials and this logarithmic K\"ahler potential is the signature of the hyperbolic geometry in supergravity. For instance, the K\"ahler metric based on the K\"ahler potential in terms of the disk variables $Z$ is given \cite{escher_in_sky} as
\begin{equation}
ds^{2}=\frac{3\alpha}{(1-ZZ^{*})^{2}}dZdZ^{*}
\end{equation}     
and the curvature of corresponding K\"ahler manifold is given by $\mathcal{R}_{\text{K\"ahler}}=-2/3\alpha$. If we decompose the field $Z$ in terms of its real field variables $Z=(x+iy)/\sqrt{3\alpha}=re^{i\theta}/\sqrt{3\alpha}$, this metric takes the form
\begin{equation}
ds^{2}=\frac{dx^{2}+dy^{2}}{\left(1-\frac{x^{2}+y^{2}}{3\alpha}\right)^{2}}=\frac{dr^{2}+r^{2}d\theta^{2}}{\left(1-\frac{r^{2}}{3\alpha}\right)^{2}}.
\end{equation}
This resembles the 2d metric of a Poincar\'e disc of the hyperbolic geometry with constant negative curvature $\mathcal{R}_{\text{Poincar\'e}}=-2/3\alpha$.  
Thus, one can claim that $\alpha$-attractor models in $\mathcal{N} = 1$ Supergravity are based on the  Poincar\'e  disc or the half-plane model of hyperbolic geometry \cite{escher_in_sky}. Based on these geometrical aspects, the observational predictions (\ref{observational_predictions}) can be read as,
\begin{equation}\label{observational_predictions_in_geometric_way}
n_{s}=1-\frac{2}{N_{e}},~~~~~~~~~~~~~r=\mathcal{R}_{\mathbb{E}}^{2}\frac{4}{N_{e}^{2}}
\end{equation}
where $\mathcal{R}_{\mathbb{E}}=\sqrt{3\alpha}$ is the radius of the Poincar\'e disc. The relation of this radius to the curvature of the Poincar\'e disc is given by $\mathcal{R}_{\mathbb{E}}^{2}=-2/\mathcal{R}_{\text{K\"ahler}}$. Thus, the primary focus of $\alpha$-attractor models is to understand the geometry of the scalar manifold from the observations rather than achieving the traditional goal of reconstruction of the inflationary potential. 

After all these success of $\alpha$-attractor models, recently some models \cite{pinhero2017non-cano-con-attra-sing-inf,pinhero2} come up with the almost identical predictions of $\alpha$-attractors and these models have a nearly flat k\"ahler geometry origin in ${\cal N}=1$ supergravity. At certain limits of these theories, these models boil down to the conformal attractor models \cite{kallosh2013universality,kallosh2013multi} which are the part of the $\alpha$-attractor family at $\alpha=1$. From this perspective, we ask whether the $\alpha$-attractor models can have a flat K\"ahler geometry origin and if such an embedding is done, whether these models are stable. We answer these questions in the affirmative. The central objective of this paper is to explicitly show that the $\alpha$-attractors can also emerge from flat K\"ahler manifolds of ${\cal N}=1$ supergravity and for any value of $\alpha>0$, these models are stable.
 
In the following section, we present two models of $\alpha$-attractors in $\mathcal{N}=1$ supergravity, each based on a logarithmic square K\"ahler potential, and we show that associated K\"ahler manifold of each of these potentials is geometrically flat. 
\section{$\alpha$-attractors from flat K\"ahler Geometry}
\subsection{Model-I}\label{model-I}
We start by defining a K\"{a}hler potential of the form 
\begin{equation}\label{kaehler_potential_log_squre_1}
K=-\frac{3\alpha}{8}\log^{2}\left[\frac{\left(1+\Phi\right)\left(1-\Phi^{*}\right)}{\left(1-\Phi\right)\left(1+\Phi^{*}\right)}\right]+SS^{*}-
\zeta (SS^*)^{2}
\end{equation}
where $\Phi$ is the inflaton superfield (non-canonical) and $S$ is the chiral multiplet which serves the job of the stabilizer field or the nilpotent superfield. During inflation, the inflaton partner $\Phi-\Phi^{*}$ and the stabilizer field attains zero vev and hence the K\"ahler potential vanishes. This is obviously due to the inflaton shift symmetry in the K\"ahler potential and this shift symmetry is explicitly visible when one writes the above K\"ahler potential in its series form as
\begin{equation}\label{eq_Kahler_potential}
K=-\left[\sum_{\substack{n=odd}}^{N}K^{(n)}\left(\Phi^{n}-\Phi^{*n}\right)\right]^2+SS^{*}-
\zeta (SS^*)^{2}
\end{equation}
with 
\begin{equation}\label{coupling_constant_eq}
K^{(n)}=\frac{1}{n}\sqrt{\frac{3\alpha}{2}},~~~~~~~~~~ N\rightarrow\infty.
\end{equation}
The index $n$ takes only odd positive integer values and $K^{(n)}$ are the dimensionless coupling constants for the self interactions of chiral superfields. This K\"ahler potential is invariant under the following shift transformation:
\begin{equation}\label{shift_symmetry}
\sum_{\substack{n=odd}}^{N}K^{(n)}\Phi^{n}\rightarrow \sum_{\substack{n=odd}}^{N}K^{(n)}\Phi^{n}+C_{N}
\end{equation}
Because of the invariance (\ref{shift_symmetry}), one can consider $\sum_{\substack{n=odd}}^{N}K^{(n)}\Phi^{n}$ as a composite field $\hat{\Phi}$, which transforms under a Nambu-Goldstone like shift symmetry. This shift symmetry generalizes the shift symmetries proposed in \cite{kawasaki2000naturalchaotic,linear_running_kinetic_inflation_takahashi,running_kinetic_inflation_takahashi} in the context of supergravity realization of the chaotic inflation and running kinetic inflation. Due to this shift symmetry, the real component of the composite field $\hat{\Phi}$ will be absent in the K\"ahler potential (\ref{eq_Kahler_potential}), and this real component $Re\left[\sum_{\substack{n=odd}}^{N}K^{(n)}\Phi^{n}\right]=Re\left[\hat{\Phi}\right]$  can be identified as the inflaton scalar field \cite{linear_running_kinetic_inflation_takahashi,running_kinetic_inflation_takahashi}. Since physics is invariant under field transformation, one can also proceed the same analysis in terms of the real non-canonical variables. In such a scenario, real part of $\Phi$ can be identified as inflaton in terms of the non-canonical chiral field $\Phi$. In order to explicitly show that the non-canonical kinetic term arising from this model also has a pole of order two with respect to the real field variable, we continue our investigation in terms of the non-canonical variables from the K\"ahler potential defined in (\ref{kaehler_potential_log_squre_1}). This investigation and the subsequent analysis based on the series K\"ahler potential (\ref{eq_Kahler_potential}) is relegated to Appendix \ref{appendix_1}. The K\"ahler potential defined in (\ref{kaehler_potential_log_squre_1})
can give the following kinetic term for inflation
\begin{multline}\label{kinetic_term_in _superfields}
\frac{1}{\sqrt{-g}}L_{kin}=-\frac{3\alpha}{\left(1-\Phi^{2}\right)\left(1-\Phi^{*2}\right)}\partial_{\mu}\Phi
\partial^{\mu}\Phi^{*}\\ -\left(1-4\zeta S^{*}S\right)\partial_{\mu}S\partial^{\mu}S^{*}
\end{multline}
As in the traditional way, we also consider a small breaking term of the shift symmetry in the superpotential,
\begin{equation}\label{superpotential1}
W=mS\Phi
\end{equation}
for the successful inflation. This small breaking term ensures a tree-level mass for the field $\Phi$ through F-term of the stabilizer field $S$. When $m\rightarrow0$, we have an enhanced shift symmetry so that our model is natural according to 't Hooft's sense \cite{t.hooft_natural}. The superpotential considered in (\ref{superpotential1}) is completely  reserved for the construction of T-model of $\alpha$- attractors. However, to construct a general $\alpha$-attractor model, one can also define a superpotential of the form,
\begin{equation}\label{general_superpotential1}
W=mSf(\Phi)
\end{equation} 
where $f(\Phi)$ is the general function of the chiral superfield $\Phi$. Now, let us decompose these complex chiral superfields $\Phi$ and $S$ into real scalar fields
\begin{equation}\label{variable_phi_chi}
\Phi=\frac{1}{\sqrt{6\alpha}}\left(\phi+i\chi\right),~~~~~~~~~S=\frac{1}{\sqrt{2}}\left(s+i\beta\right)
\end{equation} 
Based on the F-term potential of ${\cal N}=1$ SUGRA, 
\begin{equation}\label{poincare_sugra_potential}
V=e^{K}\left(D_{\Phi_{i}}WK^{ij^{*}}D_{\Phi_{j^{*}}}W^{*}-3\left|W\right|^{2}\right).
\end{equation}
and with the kinetic term (\ref{kinetic_term_in _superfields}), one will end up with the total Lagrangian for the inflation real field $\phi$ along the inflationary trajectory $\chi=s=\beta=0$ as
\begin{equation}\label{lagrangian_in_real_variables}
L=\sqrt{-g}\left[\frac{1}{2}R-\frac{1}{\left(1-\frac{\phi^{2}}{6\alpha}\right)^{2}}\frac{1}{2}\partial_{\mu}\phi\partial^{\mu}\phi-V(\phi)\right]
\end{equation}
with
\begin{equation}\label{potential1}
	V(\phi)= 
	\begin{cases}
	m^{2}\frac{\phi^{2}}{6\alpha},& \text{for    } W=mS\Phi\\
	m^{2}f^{2}(\frac{\phi}{\sqrt{6\alpha}}),& \text{for     } W=mSf(\Phi).
	\end{cases}
\end{equation}
The Lagrangian (\ref{lagrangian_in_real_variables}) with (\ref{potential1}) defines the complete $\alpha$-attractor model for inflation in terms of non-canonical real field variable $\phi$. 

Next, we study the stability of the inflationary trajectory with respect to the small fluctuations of the fields $\chi$ and $S$. In order to check this stability, one can calculate the canonical masses of all the fields orthogonal to the inflaton direction $\phi$ based on the F-term potential of SUGRA (\ref{poincare_sugra_potential}) for the theory (\ref{kaehler_potential_log_squre_1}) with (\ref{superpotential1}) or (\ref{general_superpotential1}) at $\chi=S=0$ as\footnote{It is well known that, due to the shift symmetry (\ref{shift_symmetry}) in terms of its canonical composite variables $\hat{\Phi}$, $m_{\text{Im}(\hat{\Phi})}>\mathcal{O}(H)$ for any choice of the superpotential \cite{Kallosh2010general_inflaton_pot_in_supergravity_stabilizerfield=sGoldstino}. Since physics is invariant under the field transformation, it is well expected that the canonical mass of non-canonical field $m_{\chi}>\mathcal{O}(H)$. However, we provide the explicit calculations for the same here.}
\begin{equation}\label{mass_of_chi}
m_{\chi}^{2}=6H^{2}\left(1+\epsilon-\eta/2\right)
\end{equation}
\begin{equation}\label{mass_of_S}
m_{\beta}^{2}=m_{s}^{2}=12H^{2}\left(\zeta+\epsilon/4\right)
\end{equation}
and the canonical mass of the inflaton is 
\begin{equation}\label{mass_of_inflaton}
m_{\phi}^{2}=3H^{2}\eta
\end{equation}
Here we have used $V(\phi)=3H^{2}$ where H is the Hubble's constant, and $\epsilon$ and $\eta$ are the slow-roll parameters of inflation in terms of non-canonical variable $\phi$. In order to get the canonical masses for both inflaton $\phi$ and its partner $\chi$, we have multiplied by the factor $\left(1-\phi^{2}/6\alpha\right)^{2}$ to the second derivatives of their potentials. From (\ref{mass_of_chi}) it is evident that for any value of $\alpha$, $m_{\chi}>\mathcal{O}(H)$, i.e., the field $\chi$ is heavy and reaches its minimum quickly. Moreover from (\ref{mass_of_S}), for $\zeta\geq1/12$, $m_{s}=m_{\beta}>\mathcal{O}(H)$, is also heavy and vanish. Therefore, during inflation all these fields are stabilized at the inflationary trajectory $\chi=S=0$.
\begin{figure}%[h!]
	%\centering
	%\begin{subfigure}{.5\textwidth}
	\centering
	\includegraphics[width=.9\linewidth]{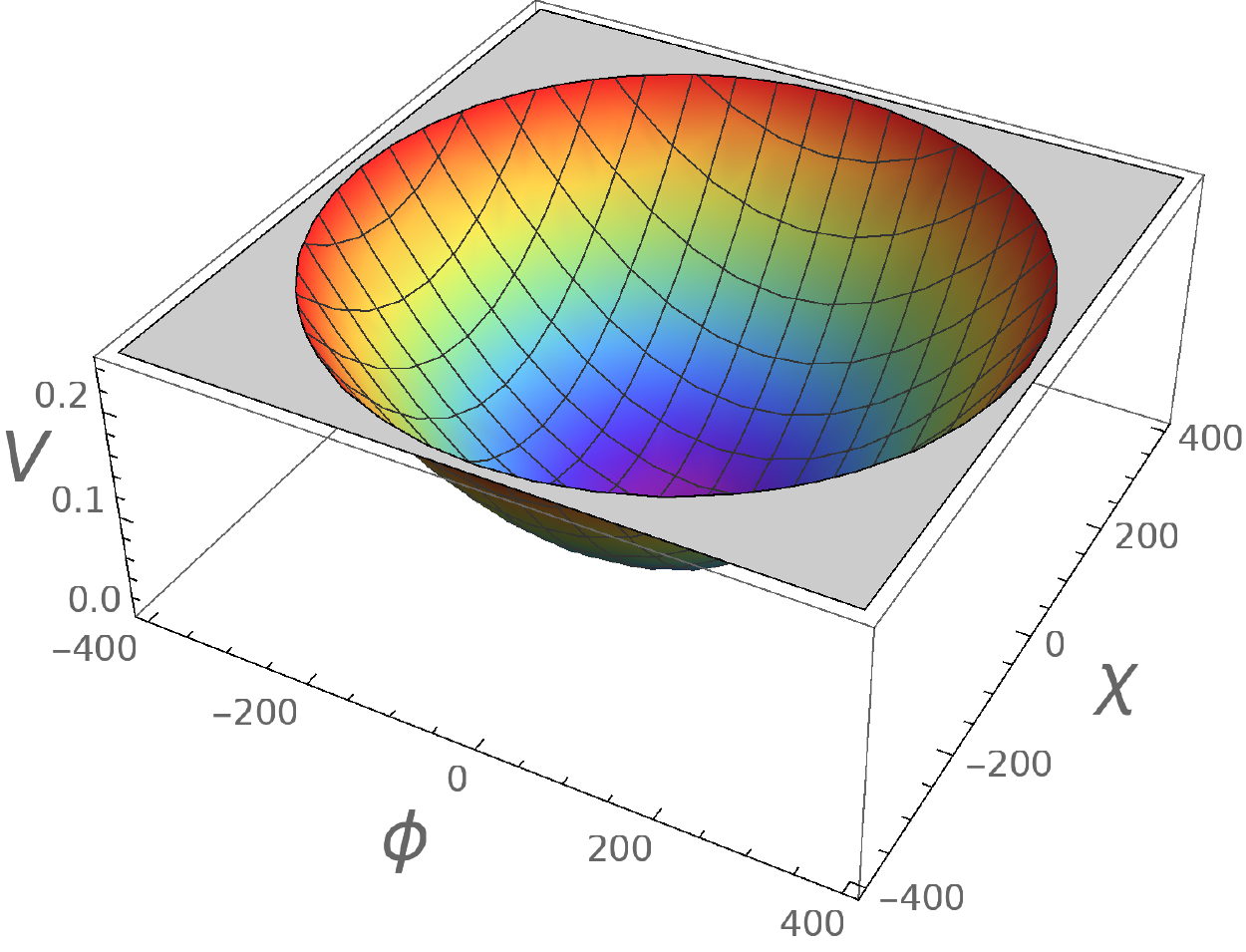}
	\caption{T-model potential for theory ({\ref{kaehler_potential_log_squre_1}}) and (\ref{superpotential1}), for the variables (\ref{variable_phi_chi}). Existence of flat direction in the potential is not visible in this set of variables.}
	\label{fig:T-model_pot_in_non-canonical_variables}
	%\end{subfigure}%
	%\begin{subfigure}{.5\textwidth}
	%	\centering
	%\includegraphics[width=.9\linewidth]{multi_field_plots_for_singlefield_canonical}
	%\caption{T-model potential, $V=\tanh^{2}\frac{\psi}{\sqrt{6}}$}
	%\label{fig:canonical T-model}
	%\end{subfigure}
	%	\caption{T-model potentials in both non-canonical Fig.(\ref{fig:non-canonical T-model}) and canonical Fig.(\ref{fig:canonical T-model}) conformal attractors scenario.}
	%\label{fig:T-model}
\end{figure} 
\begin{figure}%[h!]
	%\centering
	%\begin{subfigure}{.5\textwidth}
	\centering
	\includegraphics[width=.9\linewidth]{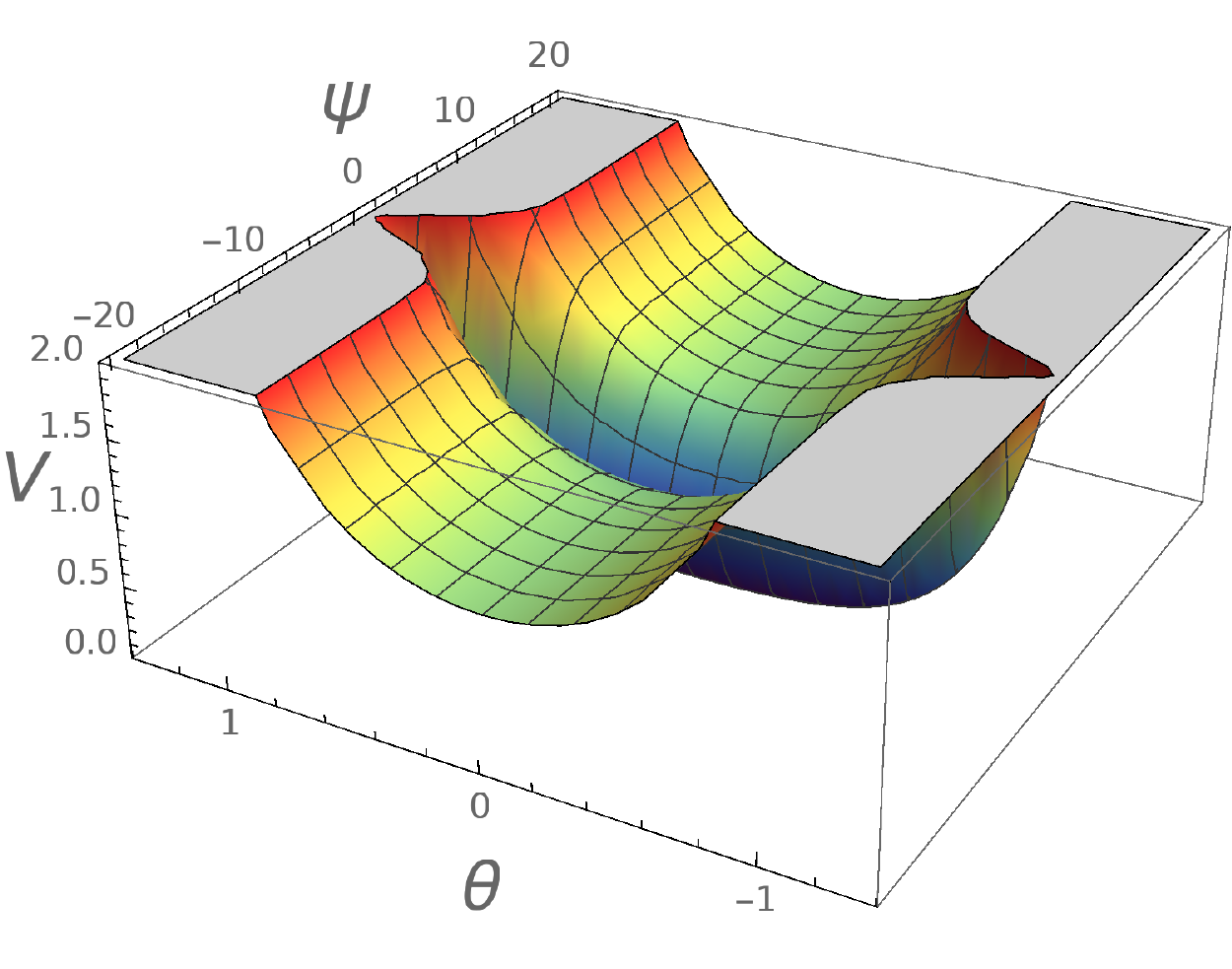}
	\caption{T-model potential for theory ({\ref{kaehler_potential_log_squre_1}}) and (\ref{superpotential1}) in terms of the canonical variables $\hat{\Phi}=(\psi+i\theta)\sqrt{6\alpha}$.}
	\label{fig:canonical_T-model_pot_with_inf_ds_valley}
	%\end{subfigure}%
	%\begin{subfigure}{.5\textwidth}
	%	\centering
	%\includegraphics[width=.9\linewidth]{multi_field_plots_for_singlefield_canonical}
	%\caption{T-model potential, $V=\tanh^{2}\frac{\psi}{\sqrt{6}}$}
	%\label{fig:canonical T-model}
	%\end{subfigure}
	%	\caption{T-model potentials in both non-canonical Fig.(\ref{fig:non-canonical T-model}) and canonical Fig.(\ref{fig:canonical T-model}) conformal attractors scenario.}
	%\label{fig:T-model}
\end{figure} 

Now, we focus on the properties of supergarvity potential (\ref{poincare_sugra_potential}) based on our model. For the simplest case, we consider T-model potential based on (\ref{kaehler_potential_log_squre_1}) and (\ref{superpotential1}) in terms of the non-canonical real variables $\Phi=(\phi+i\chi)/\sqrt{6\alpha}$. From Fig.(\ref{fig:T-model_pot_in_non-canonical_variables}), it is evident that existence of flat direction is not visible in terms of these non-canonical variables $\phi$ and $\chi$ for the theory (\ref{kaehler_potential_log_squre_1}) and (\ref{superpotential1}) due to the existence of the non-canonical kinetic term in it. In order to see the flat direction explicitly in the potential, we have to switch to more suitable variables, say, canonical variables for both $\phi$ and $\chi$ fields. So we will use the variables $\hat{\Phi}=(\psi+i\theta)\sqrt{6\alpha}$, which is related to $\Phi$ as $\Phi=\tanh\hat{\Phi}$. In terms of this new variables kinetic terms for both the fields are canonical, i.e.,
\begin{equation}\label{canonical_kin_term_for_both_fields}
L_{kin}=\frac{1}{2}\left(d\psi^{2}+d\theta^{2}\right)
\end{equation}
for any values of both $\psi$ and $\theta$. Further, the potential (\ref{poincare_sugra_potential}) at $S=0$ takes the form 
\begin{multline}\label{dS_valley_potential}
V=m^{2}e^{\frac{\theta}{\alpha}}\left|\tanh\frac{\psi+i\theta}{\sqrt{6\alpha}}\right|^{2}\\=m^{2}\frac{\cosh\sqrt{\frac{2}{3\alpha}}\psi-\cos\sqrt{\frac{2}{3\alpha}}\theta}{\cosh\sqrt{\frac{2}{3\alpha}}\psi+\cos\sqrt{\frac{2}{3\alpha}}\theta}e^{\frac{\theta^{2}}{\alpha}}
\end{multline}
This potential is depicted in the Fig.(\ref{fig:canonical_T-model_pot_with_inf_ds_valley}), which is the three dimensional plot of T-model potential with symmetric shoulders and visible flat directions. One can see that this scalar potential has a Minkowski minimum at small values of the inflaton field $\psi$ and has infinitely long dS valley at large values of $\psi$ with constant depth and width. Thus, this potential closely resembles to the one obtained from the theory based on the hyperbolic geometry of $\alpha$-attractors \cite{cosmo_attracts_nd_initial_cond_for_inflation}.

Finally, we focus on the most important part of our model, which is the geometry of the K\"ahler manifold. It is quiet evident that because of the shift symmetry (\ref{shift_symmetry}) of the K\"ahler potential (\ref{eq_Kahler_potential}), K\"ahler manifold is flat in terms of its canonical variables. More clearly, due to the shift symmetry, the series K\"ahler potential attains a canonical form in terms of its composite canonical fields $\hat{\Phi}=\sum_{\substack{n=odd}}^{N}K^{(n)}$ given by
\begin{equation}\label{kaehler_pot_in_canonical_form}
K=k\left(\hat{\Phi}-\hat{\Phi}^{*}\right)^{2}+SS^{*}-
\zeta (SS^*)^{2}
\end{equation}
where $k$ is some arbitrary constant. As an another example, one can also write the K\"ahler potential ({\ref{kaehler_potential_log_squre_1}}) in a more convenient form as
\begin{equation}
K=-\frac{3\alpha}{2}\left(\tanh^{-1}\Phi-\tanh^{-1}\Phi^{*}\right)^{2}+SS^{*}-
\zeta (SS^*)^{2}.
\end{equation}
This K\"ahler potential is also invariant under the following shift symmetric transformation:
\begin{equation}
\tanh^{-1}\Phi\rightarrow\tanh^{-1}\Phi+C
\end{equation}
In this case, we choose composite canonical field $\hat{\Phi}=\tanh^{-1}\Phi$. This is why we obtained canonical kinetic terms for both $\psi$ and $\theta$ fields with an infinite dS valley potential earlier with these composite fields (see \ref{canonical_kin_term_for_both_fields} and \ref{dS_valley_potential}). In terms of this canonical field $\hat{\Phi}$, the above K\"ahler potential takes the same form as (\ref{kaehler_pot_in_canonical_form}) which is the well known K\"ahler potential proposed in \cite{kawasaki2000naturalchaotic} with flat K\"ahler geometry. Thus, our K\"ahler manifold is also flat in terms of its canonical variables. As physics is invariant under field transformation, we expect same for non-canonical variables as well. However, we explicitly calculate the curvature for K\"ahler manifold in terms of its non-canonical variable $\Phi$ from the K\"ahler potential (\ref{kaehler_potential_log_squre_1}). Based on this K\"ahler potential, metric of the moduli space is defined as
\begin{equation}
ds^{2}=g_{\Phi\Phi^{*}}d\Phi d\Phi^{*}
\end{equation}
where
\begin{equation}
g_{\Phi\Phi^{*}}=K_{\Phi\Phi^{*}}=\frac{3\alpha}{\left(1-\Phi^{2}\right)\left(1-\Phi^{*2}\right)}.
\end{equation}
Now, we proceed to compute the non-vanishing Levi-Civita connection coefficients, Riemannian tensors, and the curvature of the moduli space associated with this K\"ahler metric. They are given as
\begin{equation}
\Gamma_{\Phi\Phi}^{\Phi}=\frac{2\Phi}{\left(1-\Phi^{2}\right)},~~~~~~~~\Gamma_{\Phi^{*}\Phi^{*}}^{\Phi^{*}}=\frac{2\Phi^{*}}{\left(1-\Phi^{*2}\right)},
\end{equation}
\begin{equation}
\mathcal{R}_{\Phi\Phi^{*}\Phi}^{\Phi}=\partial_{\Phi^{*}}\Gamma_{\Phi\Phi}^{\Phi}=0.
\end{equation}
Since all components of these Riemannian tensors vanish, curvature of the K\"ahler manifold is $\mathcal{R}_{\text{K\"ahler}}=0$. Alternatively, from the definition of curvature of K\"ahler manifold via the metric: 
\begin{equation}
\mathcal{R}_{\text{K\"ahler}}=-g_{\Phi\Phi^{*}}^{-1}\partial_{\Phi}\partial_{\Phi^{*}}\log g_{\Phi\Phi^{*}}=0.
\end{equation}
From the above, we conclude that geometry associated with our K\"ahler manifold is flat. 

\begin{figure}%[h!]
	\centering
	\includegraphics[width=.9\linewidth]{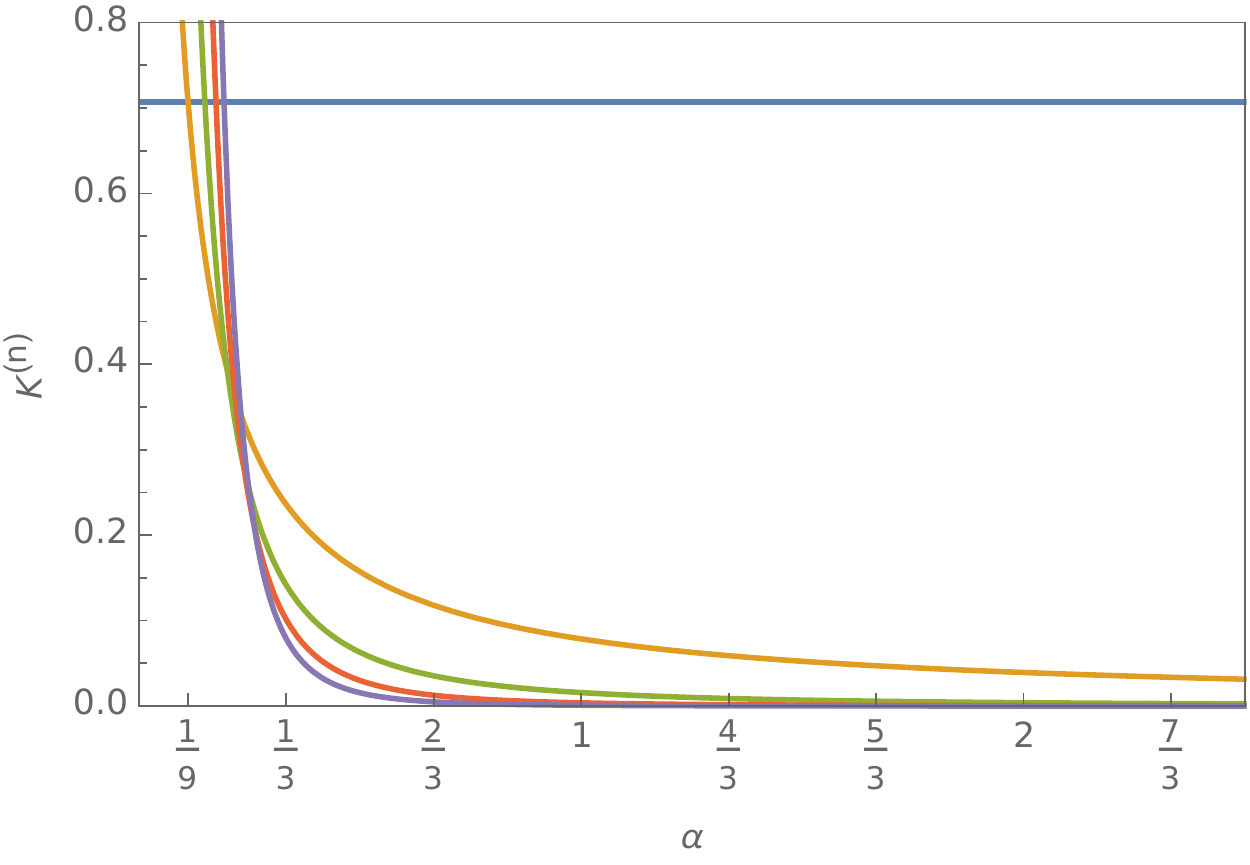}
	\caption{Behaviour of coupling constants (\ref{condition_on_coupling_constants_1}) for different values of $\alpha$ is shown. Blue, orange, green, red, and violet are stand for the values of $n=1,3,5,7~ \text{and}~ 9$ respectively.}
	\label{coupling_model1}
\end{figure}

So far, we have $\alpha$-attractor models in supergravity based on the hyperbolic K\"ahler geometry of the Poincar\'e disk or half plane with the logarithmic K\"ahler potentials. This logarithmic K\"ahler potential is the signature of the hyperbolic K\"ahler geometry of those models. In such models, parameter $\alpha$ is interpreted as the reciprocal of the curvature of the K\"ahler manifold. In contrast, we present an $\alpha$-attractor model based on the square of the logarithmic K\"ahler potential with a vanishing curvature. However, this leaves us with the following question - how can we interpret the parameter $\alpha$? Without going into the further details of K\"ahler geometry, from ({\ref{coupling_constant_eq}}), a possible interpretation is that the parameter $\alpha$ plays the roll of coupling constants, which represents the self interaction of the inflaton superfields. For a non-canonically normalized field $\Phi=(\phi+i\chi)/\sqrt{2}$, (\ref{coupling_constant_eq}) can be modified as 
\begin{equation}\label{condition_on_coupling_constants_1}
K^{(n)}=\frac{1}{(6\alpha)^{\frac{n-1}{2}}}\frac{\left(\sqrt{2}\right)^{n-2}}{n}
\end{equation}
Bahaviour of these coupling constants under different values of $\alpha$ is shown in Fig.(\ref{coupling_model1}).

\subsection{Model-II}\label{model-II}
\begin{figure}%[h!]
	%\centering
	%\begin{subfigure}{.5\textwidth}
	\centering
	\includegraphics[width=.9\linewidth]{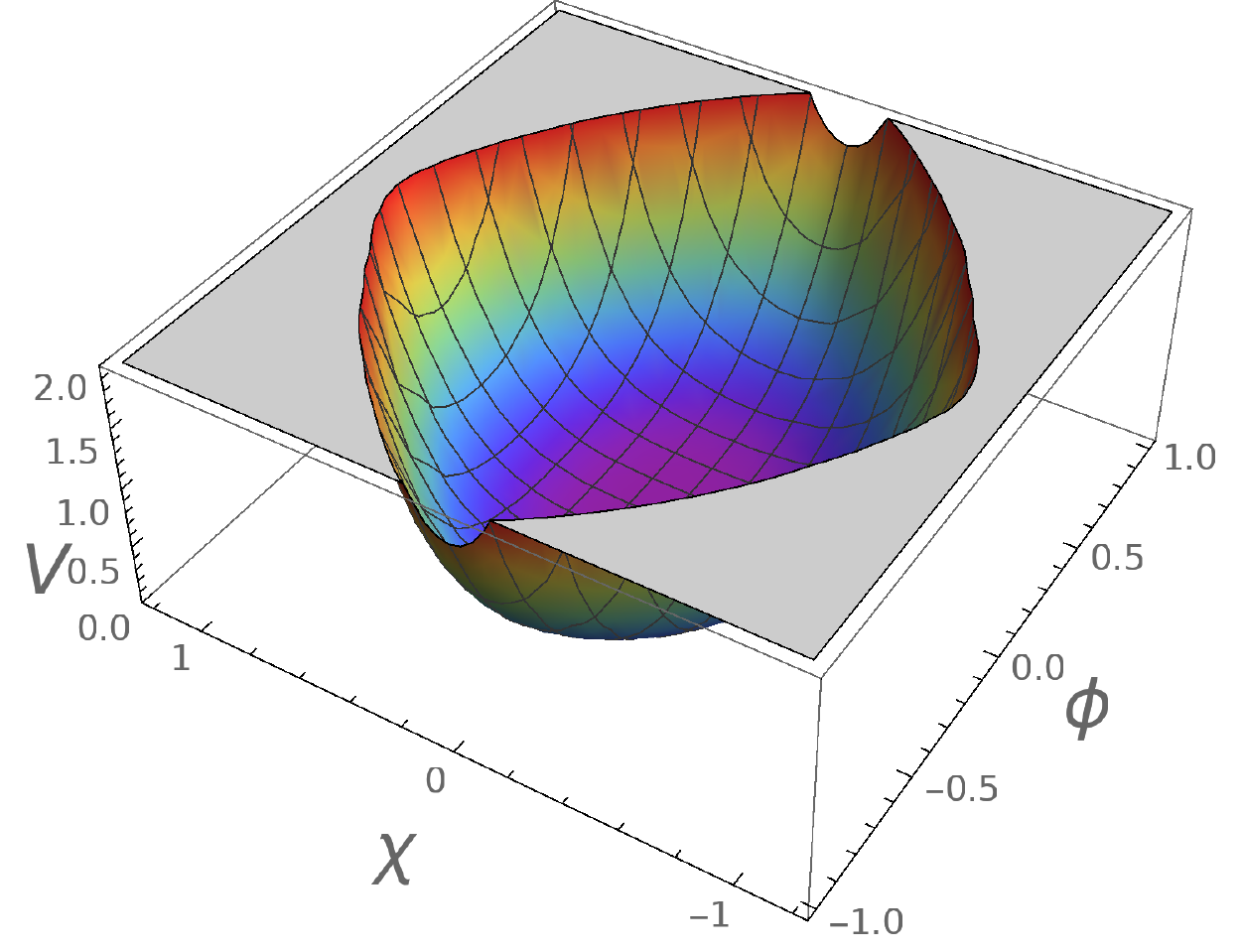}
	\caption{T-model potential for the theory (\ref{kaehler_potential_log_squre_2}) and (\ref{superpotential_t-model_2}) in terms of non-canonical varibles $\phi$ and $\chi$ with non-visible flat directions.} 
	\label{model_2_fig}
	%\end{subfigure}%
	%\begin{subfigure}{.5\textwidth}
	%	\centering
	%\includegraphics[width=.9\linewidth]{multi_field_plots_for_singlefield_canonical}
	%\caption{T-model potential, $V=\tanh^{2}\frac{\psi}{\sqrt{6}}$}
	%\label{fig:canonical T-model}
	%\end{subfigure}
	%	\caption{T-model potentials in both non-canonical Fig.(\ref{fig:non-canonical T-model}) and canonical Fig.(\ref{fig:canonical T-model}) conformal attractors scenario.}
	%\label{fig:T-model}
\end{figure}

In this section, we consider the K\"ahler potential for the chiral superfield $\Phi$ as
\begin{equation}\label{kaehler_potential_log_squre_2}
K=-\frac{3\alpha}{8}\log^{2}\left[\frac{\left(1+\Phi^{2}\right)\left(1-\Phi^{*2}\right)}{\left(1-\Phi^{2}\right)\left(1+\Phi^{*2}\right)}\right]+SS^{*}-
\zeta (SS^*)^{2}
\end{equation}
This K\"ahler potential also has a shift symmetry in terms of its canonical composite chiral fields given by $\hat{\Phi}=\sum_{\substack{n=odd}}^{N}K^{(2n)}\Phi^{2n}$. More explicitly, we write the above K\"ahler potential in the following form:
\begin{multline}\label{eq_Kahler_potential_2}
K=-\left[\sum_{\substack{n=odd}}^{N}\frac{K^{(2n)}
}{M_{pl}^{2n-2}}\left(\Phi^{2n}-\Phi^{*2n}\right)\right]^2\\+SS^{*}-
\zeta (SS^*)^{2}
\end{multline}
which is invariant under the following transformation
\begin{equation}
\sum_{\substack{n=odd}}^{N}K^{(2n)}\Phi^{2n}\rightarrow \sum_{\substack{n=odd}}^{N}K^{(2n)}\Phi^{2n}+C_{N}.
\end{equation}
We also consider a shift symmetry breaking superpotential
\begin{equation}\label{superpotential_t-model_2}
W=\sqrt{\lambda} S\Phi^{2}
\end{equation}
or generally,
\begin{equation}\label{general_superpotential_2}
W=Sf(\Phi^{2}).
\end{equation}
One may see that this model is related to the previous model (\ref{eq_Kahler_potential}) and (\ref{superpotential1}) (or (\ref{general_superpotential1})) by a scaled transformation in the field $\Phi\rightarrow\Phi^{2}$. Since this scaling is non-linear and these models are not related by any K\"ahler transformations, one can consider these as two different models. The K\"ahler potential defined in (\ref{kaehler_potential_log_squre_2}) produces the following kinetic term as
\begin{multline}\label{kinetic_term_in_superfields_2}
\frac{1}{\sqrt{-g}}L_{kin}=-\frac{12\alpha}{\left(1-\Phi^{4}\right)\left(1-\Phi^{*4}\right)}\partial_{\mu}\Phi
\partial^{\mu}\Phi^{*}\\ -\left(1-4\zeta S^{*}S\right)\partial_{\mu}S\partial^{\mu}S^{*}
\end{multline}
decomposing these complex variables into real variables as follows
\begin{equation}\label{in_terms_of_real_variables}
\Phi=\frac{1}{(6\alpha)^{\frac{1}{4}}}\left(\phi+i\chi\right),~~~~~~~~~S=\frac{1}{\sqrt{2}}\left(s+i\beta\right).
\end{equation} 
With the help of (\ref{poincare_sugra_potential}), for the theory (\ref{kinetic_term_in_superfields_2}) and (\ref{superpotential_t-model_2}) (or \ref{general_superpotential_2}), one can write the total Lagrangian for inflation along $S=\chi=0$ as
\begin{equation}\label{lagrangian_in_real_variables_2}
L=\sqrt{-g}\left[\frac{1}{2}R-\frac{4\phi^{2}}{\left(1-\frac{\phi^{4}}{6\alpha}\right)^{2}}\frac{1}{2}\partial_{\mu}\phi\partial^{\mu}\phi-V(\phi)\right]
\end{equation}
where
\begin{equation}\label{potential_model2}
V= 
\begin{cases}
\frac{\lambda}{6\alpha}\phi^4,& \text{for    } W=\sqrt{\lambda}S\Phi\\
f^{2}\left(\frac{\phi^{2}}{\sqrt{6\alpha}}\right),& \text{for     } W=Sf(\Phi^{2}).
\end{cases}
\end{equation}
Note that the kinetic term has four poles of order two at $((6\alpha)^{\frac{1}{4}},-(6\alpha)^{\frac{1}{4}},i(6\alpha)^{\frac{1}{4}},-i(6\alpha)^{\frac{1}{4}})$ and has a leading pole of order two with residue $\frac{3}{2}\alpha$. This is the required condition in kinetic term for the $\alpha$-attractors in its non-canonical variables so that the observational predictions are to a large extent determined by this term, rather than by the potential. In order to switch to the canonical variable, we solve the equation
\begin{equation}
\frac{2\phi}{\left(1-\frac{\phi^{2}}{6\alpha}\right)}\partial\phi=\partial\psi
\end{equation}
which gives
\begin{equation}
\phi=\left(6\alpha\right)^{\frac{1}{4}}\tanh^{\frac{1}{2}}\left(\frac{\psi}{\sqrt{6\alpha}}\right)
\end{equation}
Thus, the total Lagrangian for the $\alpha$-attractor model in terms of canonical variable $\psi$ reads
\begin{equation}
L=\sqrt{-g}\left[\frac{1}{2}R-\frac{1}{2}\partial_{\mu}\psi\partial^{\mu}\psi-f^{2}\left(\tanh\left(\frac{\psi}{\sqrt{6\alpha}}\right)\right)\right].
\end{equation}
Masses of all stabilized heavy fields along inflationary trajectory $S=\chi=0$ are given by
\begin{equation}\label{mass_of_chi_model2}
m_{\chi}^{2}=6H^{2}\left(1-\frac{3}{4}\epsilon+\eta/2\right)
\end{equation}
and
\begin{equation}\label{mass_of_S_model2}
m_{\beta}^{2}=m_{s}^{2}=12H^{2}\left(\zeta+\epsilon/8\right).
\end{equation}
These masses are slightly different compared to the masses of Model-I. The scalar potential for the model (\ref{eq_Kahler_potential_2}) and (\ref{superpotential_t-model_2}) in terms of the variables (\ref{in_terms_of_real_variables}) is shown in Fig.(\ref{model_2_fig}). As in the earlier case, here also the flat directions are not visible in terms of these variables. Migrating to more adequate variables $\Phi=\tanh^{1/2}\hat{\Phi}$ with $\hat{\Phi}=(\psi+i\theta)\sqrt{6\alpha}$, we get exactly the same potential (\ref{dS_valley_potential}) that we obtained in the previous model, which has visible flat directions. It is shown in Fig.(\ref{fig:canonical_T-model_pot_with_inf_ds_valley}). Moreover, the kinetic term is canonical for inflaton and its partner (see \ref{canonical_kin_term_for_both_fields}).

\begin{figure}%[h!]
	\centering
	\includegraphics[width=.9\linewidth]{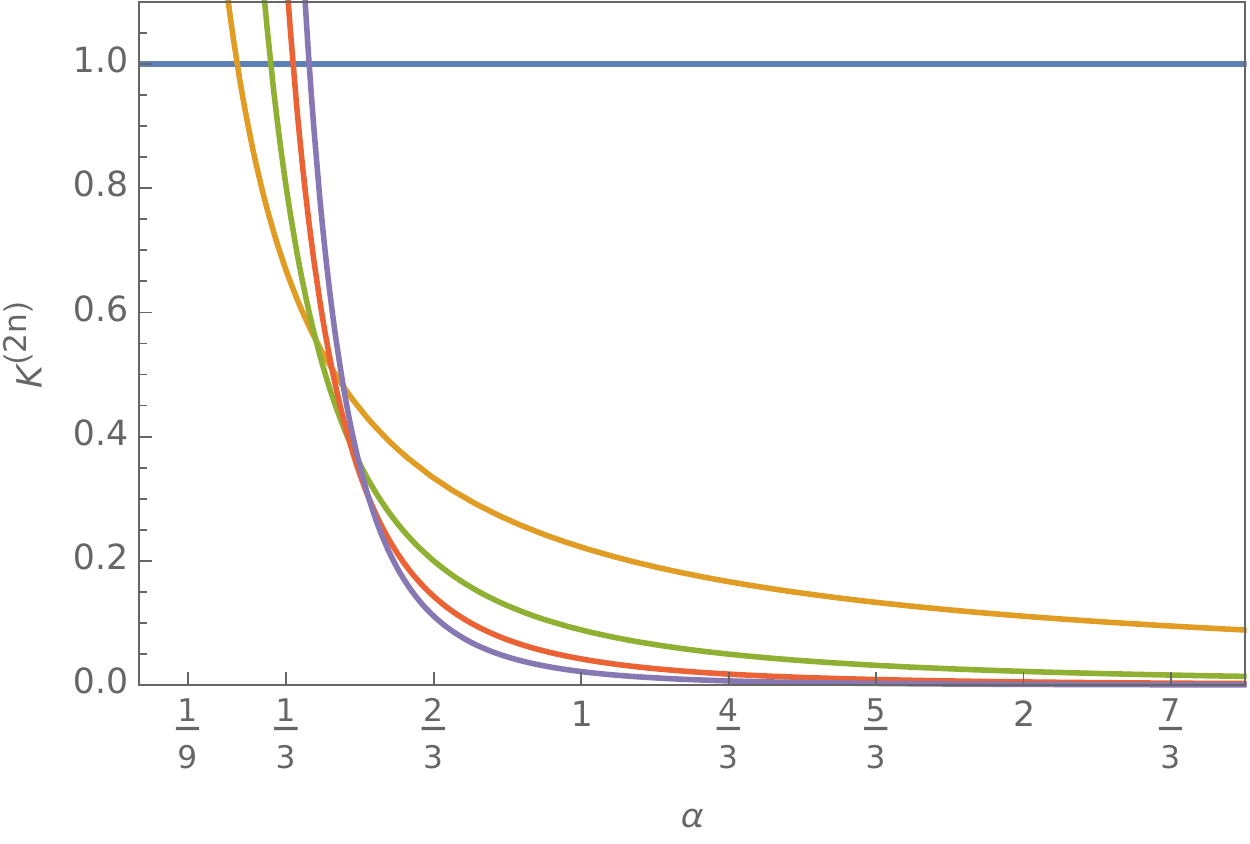}
	\caption{ Behaviour of coupling constants (\ref{coupling_constants2}) for different values of $\alpha$ is shown. Blue, orange, green, red, and violet are stand for the values of $n=1,3,5,7~ \text{and}~ 9$ respectively.}
	\label{fig:coupling constants2}
\end{figure}

Finally, the metric of the moduli space obtained from the K\"ahler potential (\ref{kaehler_potential_log_squre_2}),
\begin{equation}
ds^{2}=\frac{12\alpha}{\left(1-\Phi^{4}\right)\left(1-\Phi^{*4}\right)}d\Phi d\Phi^{*}
\end{equation}
gives non-vanishing connection coefficients as follows:
\begin{equation}
\Gamma_{\Phi\Phi}^{\Phi}=\frac{1+3\Phi^{4}}{\Phi\left(1-\Phi^{4}\right)},~~~~~~~~\Gamma_{\Phi^{*}\Phi^{*}}^{\Phi^{*}}=\frac{1+3\Phi^{*4}}{\Phi^{*}\left(1-\Phi^{*4}\right)}.
\end{equation}
However, all components of Riemannian tensors vanish and hence, the geometry of the K\"ahler manifold is flat. As before, the parameter $\alpha$ here can be related to the coupling constants as
\begin{equation}\label{coupling_constants2}
K^{(2n)}=\frac{2^{n-1}}{n}\frac{1}{(6\alpha)^{\frac{n-1}{2}}}
\end{equation}
and its behaviour is shown in Fig.(\ref{fig:coupling constants2}).
\section{Concluding remarks}
We have presented a couple of $\alpha$-attractors models based on $\mathcal{N}=1$  supergravity with the use of logarithmic square K\"ahler potentials. The associated geometry of such K\"ahler manifolds have been found to be flat where all components of Riemannian tensors vanish. The masses of all fields which are orthogonal to inflaton direction have been stabilized during inflation. The scalar potential $V$ in its canonical form seems to have an approximately similar kind of behavior compared to the potential which is obtained from the theory based on the hyperbolic geometry of the $\alpha$-attractors. 

In contrast to the supergravity $\alpha$-attractor models based on the hyperbolic geometry in the literature, we have explicitly shown that one can obtain the same $\alpha$-attractors from the flat K\"ahler manifolds. Therefore, it is evident that cosmological $\alpha$-attractor models can emerge not only from the hyperbolic K\"ahler manifolds but also from the geometry of flat K\"ahler manifolds.

%%%%%%%%%%%%%%%%%%%%%%%%%%%%%%%%%%%%%%%%%%%%%%%%%%%%%%%%%%%%%%%%%%%%%%%%%%%%%%%%%%%%%%%%%%%%%%%%%%%%%%%%%%%%%%%%%%%%%%%%%%%%%%%%%%%%%%%%%%%%%%%%%%%%%%%%%%%%%%%%%%%%%%%%%%%%%%%%%%%%%%%%%%%%%%%%%%%%%%%%%%%%%%%%%%%%%%%%%%%%%
%%%%%%%%%%%%%%%%%%%%%%%%%%%%%%%%%%%%%%%%%%%%%%%%%%%%%%%%%%%%%%%%%%%%%%%%%%%%%%%%%%%%%%%%%%%%%%%%%%%%%%%%%%%%%%%%%%%%%%%%%%%%%%%%%%%%%%%%%%%%%%%%%%%%%%%%%%%%%%%%%%%%%%%%%%%%%%%%%%%%%%%%%%%%%%%%%%%%%%%%

\section*{Acknowledgments}

We would like to thank Supratik Pal, Arindam Chatterjee,  Shubhabrata Das, Debabrata Chandra, Abhishek Naskar and Arnab Paul for the enlightening discussions. T.P gratefully acknowledge the support from Senior Research fellowship (Order No. DS/18-19/0616) of the Indian Statistical Institute (ISI), Kolkata.
%%%%%%%%%%%%%%%%%%%%%%%%%%%%%%%%%%%%%%%%%%%%%%%%%%%%%%%%%%%%%%%%%%%%%%%%%%%%%%%%%%%%%%%%%%%%%%%%%%%%%%%%%%%%%%%%%%%%%%%%%%%%%%%%%%%%%%%%%%%%%%%%%%%%%%%%%%%%%%%%%%%%%%%%%%%%%%%%%%%%%%%%
%%%%%%%%%%%%%%%%%%%%%%%%%%%%%%%%%%%%%%%%%%%%%%%%%%%%%%%%%%%%%%%%%%%%%%%%%%%%%%%%%%%%%%%%%%%%%%%%%%%%%%%%%%%%%%%%%%%%%%%%%%%%%%%%%%%%%%%%%%%%%%%%%%%%%%%%%%%%%%%%%%%%%%%%%%%%%%%%%%%%%%%%%%%%%%%%%%%

\appendix
\section{Derivation of Model-I in series form}\label{appendix_1}
The K\"ahler potential defined in (\ref{eq_Kahler_potential})
can give the following kinetic term for inflation as
\begin{multline}\label{eq_kinetic_term_in_superfields}
\frac{1}{\sqrt{-g}}L_{kin}=-2\left[\sum_{\substack{n=odd}}^{N}K^{(n)}
n\Phi^{n-1}\sum_{\substack{n=odd}}^{N}K^{(n)}
n\Phi^{*n-1}\right]\\
\times \partial_{\mu}\Phi_{n}
\partial^{\mu}\Phi^{*}_{n}-\left(1-4\zeta S^{*}S\right)\partial_{\mu}S\partial^{\mu}S^{*}
\end{multline}
with
\begin{equation}\label{condition_on_coupling_constants}
K^{(n)}=\frac{1}{(6\alpha)^{\frac{n-1}{2}}}\frac{\left(\sqrt{2}\right)^{n-2}}{n}
\end{equation}
 solve the equation for the canonical normalization of the form,
\begin{equation}
\sqrt{2}\sum_{\substack{n=\text{odd}}}^{N}K^{(n)}n\Phi^{n-1}\partial\Phi=\partial\hat{\Phi}
\end{equation}
which yields,
\begin{equation}\label{canonical_normalization_solution}
\sum_{\substack{n=\text{odd}}}^{N}\frac{\left(\sqrt{2}\right)^{n-1}
}{n(6\alpha)^{\frac{n-1}{2}}}\Phi^{n}=\hat{\Phi}
\end{equation}
Here the values of $K^{(n)}$ are adopted from (\ref{condition_on_coupling_constants}). Note that, Left hand side of the above (\ref{canonical_normalization_solution}) represents the series expansion of $\sqrt{3\alpha}\tanh^{-1}\left(\frac{\Phi}{\sqrt{3\alpha}}\right)$ for large values of $N$. Hence the chiral field $\Phi$ can be represented in terms of canonically normalized field $\hat{\Phi}$ as,
\begin{equation}\label{Phi_in_canonical_cariable_phicap}
\Phi=\sqrt{3\alpha}\tanh\left(\frac{\hat{\Phi}}{\sqrt{3\alpha}}\right)
\end{equation}
Thus in terms of canonical fields total Lagrangian for inflation takes the form, 
\begin{equation}
L=\sqrt{-g}\left[\frac{1}{2}R-\partial_{\mu}\hat{\Phi}\partial^{\mu}\hat{\Phi^{*}}-3\alpha m^{2}\left |\tanh\left(\frac{\hat{\Phi}}{\sqrt{3\alpha}}\right)  \right |^{2}\right]
\end{equation}
Decomposing $\hat{\Phi}$ into real and imaginary parts as $\hat{\Phi}=\frac{1}{\sqrt{2}}\left(\hat{\psi}+i\hat{\chi}\right)$ and at $\chi=o$ Lagrangian for inflation in terms of the real field $\hat{\psi}$ as, 
\begin{equation}
L=\sqrt{-g}\left[\frac{1}{2}R-\frac{1}{2}\partial_{\mu}\hat{\psi}\partial^{\mu}\hat{\psi}-3\alpha m^{2}\tanh^{2}\left(\frac{\hat{\psi}}{\sqrt{6\alpha}}\right)\right]
\end{equation}
Which defines the T-model of $\alpha$-attractors. For the general case, or for the superpotential of (\ref{general_superpotential1}), Lagrangian reads the form by using (\ref{Phi_in_canonical_cariable_phicap}) as,
\begin{equation}\label{general_alpha_attract_in_canonical_psicap}
L=\sqrt{-g}\left[\frac{1}{2}R-\frac{1}{2}\partial_{\mu}\hat{\psi}\partial^{\mu}\hat{\psi}-f^{2}\left(\sqrt{3\alpha}\tanh\left(\frac{\hat{\psi}}{\sqrt{6\alpha}}\right)\right)\right]
\end{equation}
This (\ref{general_alpha_attract_in_canonical_psicap}) defines the $\alpha$-attractor Lagrangian in terms of the real canonical variable $\hat{\psi}$. However physics is invariant under field transformation, one can also derive the above same Lagrangian in terms of the real non-canonical variable. So, in such a scenario, real part of $\Phi$ can be identified as inflaton in terms of the non-canonical chiral field $\Phi$. By decomposing $\Phi$ in terms of real and imaginary components $\Phi=\frac{1}{\sqrt{2}}\left(\psi+i\chi\right)$, and by assuming along flat direction $S=\chi=0$, the (\ref{eq_kinetic_term_in_superfields}) can be written as 
\begin{equation}\label{kinetic_term_series_final}
\frac{1}{\sqrt{-g}}L_{kin}=-\left[\sum_{\substack{n=odd}}^{N}\frac{1}{\left(6\alpha\right)^{\frac{n-1}{2}}}\phi^{n-1}\right]^{2}\frac{1}{2}\partial_{\mu}\phi\partial^{\mu}\phi
\end{equation}
For large values of $N$, and for the field $\phi$ satisfies the condition $\phi^{2}<6\alpha$, one can identifies the series defined in the (\ref{kinetic_term_series_final})
is a the Taylor's series expansion of the term $\frac{1}{\left(1-\frac{\phi^{2}}{6\alpha}\right)^{2}}$. Thus the final kinetic term is
\begin{equation}\label{final_non_canonical_kin_term_in_real_field}
\frac{1}{\sqrt{-g}}L_{kin}=-\frac{1}{\left(1-\frac{\phi^{2}}{6\alpha}\right)^{2}}\frac{1}{2}\partial_{\mu}\phi\partial^{\mu}\phi
\end{equation}
Now for the superpotential defined in (\ref{superpotential1}) or for (\ref{general_superpotential1}), one can write the potential in terms of real field $\phi$ for $S=\chi=0$ as,
\begin{equation}\label{final_T-potential}
V=\frac{m^{2}}{2}\phi^{2}
\end{equation}
or 
\begin{equation}\label{final_general_potential}
V=f^{2}\left(\frac{\phi}{\sqrt{2}}\right)
\end{equation}
respectively. Clubbing together (\ref{final_non_canonical_kin_term_in_real_field}) and (\ref{final_T-potential}) we get the T-model Lagarngian in real canonical variable as
\begin{equation}
L=\sqrt{-g}\left[\frac{1}{2}R-\frac{1}{\left(1-\frac{\phi^{2}}{6\alpha}\right)^{2}}\frac{1}{2}\partial_{\mu}\phi\partial^{\mu}\phi-\frac{m^{2}}{2}\phi^{2}\right]
\end{equation}
or, generally
\begin{equation}
L=\sqrt{-g}\left[\frac{1}{2}R-\frac{1}{\left(1-\frac{\phi^{2}}{6\alpha}\right)^{2}}\frac{1}{2}\partial_{\mu}\phi\partial^{\mu}\phi-f^{2}\left(\frac{\phi}{\sqrt{2}}\right)\right]
\end{equation}
Which defines the $\alpha$-attractor Lagrangian before switching to canonical variable.
	
\section{Derivation of Model-II in series form}\label{appendix_2}
Kinetic term for the K\"ahler potential (\ref{eq_Kahler_potential_2}) is
\begin{multline}\label{eq_kinetic_term_in_series_superfields_2}
\frac{1}{\sqrt{-g}}L_{kin}=-2\left[\sum_{\substack{n=odd}}K^{(2n)}
2n\Phi^{2n-1}\right. \\ \left. \times\sum_{\substack{n=odd}}K^{(2n)}
2n\Phi^{*2n-1}\right]
\partial_{\mu}\Phi_{n}
\partial^{\mu}\Phi^{*}_{n}\\-\left(1-4\zeta S^{*}S\right)\partial_{\mu}S\partial^{\mu}S^{*}
\end{multline}
with (\ref{coupling_constants2}).
Decomposing the field $\Phi$ into real and imaginary parts as $\Phi=\frac{1}{\sqrt{2}}\left(\psi+i\chi\right)$ and considering along the flat direction $X=\chi=0$, we get
\begin{equation}\label{kinetic_term_in decomposed_fields_with_fine_tuned_Ks}
\frac{1}{\sqrt{-g}}L_{kin}=-\left[\sum_{\substack{n=odd}}\frac{2}{(6\alpha)^{\frac{n-1}{2}}}\phi^{2n-1}\right]^{2}\frac{1}{2}\partial_{\mu}\phi\partial^{\mu}\phi
\end{equation}
This can be explicitly write in the form,
\begin{multline}\label{kin_series_form_expanded}
\frac{1}{\sqrt{-g}}L_{kin}=\left(1+\frac{1}{6\alpha}\phi^{4}+\frac{1}{(6\alpha)^{2}}\phi^{8}+\frac{1}{(6\alpha)^{3}}\phi^{12}+\dots
\right)^{2}\\\times-\frac{4\phi^{2}}{2}\partial_{\mu}\phi\partial^{\mu}\phi
\end{multline}
Now for very large values of $N$ and if the field $\phi$ satisfies the condition $\phi^{4}<6\alpha$, one can recognize the series defined in (\ref{kin_series_form_expanded}) as the coefficient of $\frac{4\phi^{2}}{2}\partial_{\mu}\phi\partial^{\mu}\phi$ is as the Taylor's series expansion of the term $\frac{1}{\left(1-\frac{\phi^{4}}{6\alpha}\right)^{2}}$. Thus the final kinetic term in terms of real non-canonical variable takes the form
\begin{equation}\label{kinetic_term_final2}
\frac{1}{\sqrt{-g}}L_{kin}=-\frac{4\phi^{2}}{\left(1-\frac{\phi^{4}}{6\alpha}\right)^{2}}\frac{1}{2}\partial_{\mu}\phi\partial^{\mu}\phi
\end{equation}
and the scalar potential reads
\begin{equation}\label{potential2}
V= 
\begin{cases}
\frac{\lambda}{4}\phi^4,& \text{for    } W=\sqrt{\lambda}S\Phi\\
f^{2}\left(\phi^{2}\right),& \text{for     } W=Sf(\Phi^{2})
\end{cases}
\end{equation}
Clubbing (\ref{kinetic_term_final2}) and (\ref{potential2}) together, we get the total Lagrangian for inflation in non-canonical real variable as,
\begin{equation}
L=\sqrt{-g}\left[\frac{1}{2}R-\frac{4\phi^{2}}{\left(1-\frac{\phi^{2}}{6\alpha}\right)^{2}}\frac{1}{2}\partial_{\mu}\phi\partial^{\mu}\phi-f^{2}\left(\phi^{2}\right)\right]
\end{equation}
Solve the equation for canonical normalization
\begin{equation}
\frac{2\phi}{\left(1-\frac{\phi^{2}}{6\alpha}\right)}\partial\phi=\partial\psi
\end{equation}
one yields,
\begin{equation}
\phi=\left(6\alpha\right)^{\frac{1}{4}}\tanh^{\frac{1}{2}}\left(\frac{\psi}{\sqrt{6\alpha}}\right)
\end{equation}
So, finally in terms if this canonical field $\psi$, the Lagrangian reads
\begin{equation}
L=\sqrt{-g}\left[\frac{1}{2}R-\frac{1}{2}\partial_{\mu}\psi\partial^{\mu}\psi-f^{2}\left(\sqrt{\frac{3\alpha}{2}}\tanh\left(\frac{\psi}{\sqrt{6\alpha}}\right)\right)\right]
\end{equation}
Which defines total Lagrangian for $\alpha$-attractors.

%%%%%%%%%%%%%%%%%%%%%%%%%%%%%%%%%%%%%%%%%%%%%%%%%%%%%%%%%%%%%%%%%%%%%%%%%%%%%%%%%%%%%%%%%%%%%%%%%%%%%%%%%%%%%%%%%%%%%%%%%%%%%%%%%%%%%%%%%%%%%%%%%%%%%%%%%%%%%%%%%%%%%%%%%%%

\end{document}